\documentclass[a4paper,11pt]{article}
\pdfoutput=1 

\usepackage{jheppub} 

\usepackage[T1]{fontenc} 
\usepackage{amssymb}
\usepackage{amsfonts}

\newcommand{\rp}{\right)}
\newcommand{\lp}{\left(}

\newcommand{\bra}{\right>}
\newcommand{\ket}{\left<}

\newcommand{\ra}{\rightarrow}

\newcommand{\pr}{\partial}

\newcommand{\gag}{\quad \mbox{and} \quad}
\newcommand{\gwg}{\quad \mbox{where} \quad}

\newcommand{\vx}{{\vec{x}}}
\newcommand{\vy}{{\vec{y}}}
\newcommand{\vk}{{\vec{k}}}
\newcommand{\ak}{{a_{\vec{k}}}}
\newcommand{\adk}{{a_{\vec{k}}^\dagger}}
\newcommand{\bk}{{b_{\vec{k}}}}

\newcommand{\gbb}{{G_{BB}}}
\newcommand{\gio}{{G_{in-out}}}

\newcommand{\obs}{{\mathcal{O}}}

\renewcommand{\Im}{\operatorname{Im}}
\renewcommand{\Re}{\operatorname{Re}}

\newcommand{\A}{\mathcal{A}}

\title{\boldmath Gravitational Particle Production and Locality of the Wavefunction-Squared}

 \author{Emine Şeyma Kutluk}
 \affiliation{Scuola Normale Superiore,\\Piazza dei Cavalieri 56121,\\ Pisa PI Italy}

\emailAdd{emine.kutluk@sns.it}

\abstract{We show that no particle production due to the background geometry implies the locality of the wavefunction-squared for free heavy masses on expanding cosmologies. On global de Sitter of odd dimensions these two stem from the same identity, whereas in the Poincare patch neither occur under any condition. These two phenomena both depend on the behaviour of the out modes at far future and we show that they almost always come together for systems with an out mode that has a phase with separated time and momentum dependency at far future. This will be true for most expanding cosmologies. We comment on how the discussion can be connected to Green's functions and on some previous results in the literature that are seemingly contrary to our conclusions.}

\begin{document} 
\maketitle
\flushbottom

\section{Introduction}
\label{sec:intro}

Cosmological correlators, the quantum mechanical expectation values of perturbative fields in the early universe are our probes to the formation of the universe. We have access to these quantities through the cosmic microwave background radiation measurements, and efforts to produce theoretical predictions of them beyond the leading Gaussian order began with \cite{maldacena-non-gaussian}. The bootstrapping techniques accelerated the progress in the field and now many different shapes of correlations are connected to the underlying physics, see e.g. \cite{snowmass} for a review.

Different methodologies to calculate these quantities can be divided into three categories as in-in formalism, wavefunction method and bootstrap method \cite{baumann}. The wavefunction \footnote{Which is a functional for a field theory.} is the probability of a certain field configuration to be realized at a certain time, and a cosmological correlator can be written as a path integral of the related fields at that instance weighted with the wavefunction-squared--the modulus squared of the wavefunction. The inflationary correlators are typically calculated at the end of the inflationary phase, or when the deviation from the exact de Sitter can be omitted--at the far future of de Sitter. 

This wavefunction at the boundary of de Sitter can be related to a possible CFT dual. This was discussed in \cite{maldacena-non-gaussian}, where it has been noted that the wavefunction-squared of a free massless field that has started off in the Bunch-Davies vacuum at the far past of the Poincare patch of de Sitter is local at the late time surface when the spacetime dimension is odd.\footnote{See also \cite{behrang} for other theories with local wavefunction-squared beyond the free theory.} (Throughout the paper, when we say dimension without further specification we will always mean the spacetime dimension.)

Independently of this, it was noted in \cite{strominger-conformal} that on odd dimensional global de Sitter ``there is no particle production'' of heavy fields. This means that starting off in a properly defined $in$ state at the far past, there will be no $out$ state particles at the far future. Here the $in$ and the $out$ states are vacuum states that are chosen to have oscillating behaviour at the far past and the far future respectively. The no-particle-production result simply states that the $in$ and the $out$ states are the same in odd dimensional global de Sitter. This discrepancy of behaviour in between the odd and even dimensions for both phenomena raises the question of whether they are connected. 

The situation is very reminiscent of the wave-propagation in flat space: for this case it is known that in even dimensional spacetime wave propagates without a tail, or in other words follows Huygens' principle \cite{Hassani,Dai:2013cwa}. This can be expressed in terms of the retarded Green's functions, where the absence of the tail implies that the retarded Green's function is a Dirac delta or a sum of its derivatives, namely
\begin{equation}
G_R^{(2n+4)}(t,\vx,t',\vx') \sim \theta(t-t')\lp \frac{1}{r} \frac{d}{dr} \rp^n \frac{\delta(t-t'-r)}{r} \gwg r:= |\vx-\vx'| \ 
\end{equation}
and $2n+4$ is the dimension of the spacetime. For de Sitter the situation is more complicated, except for the conformal masses, as Green's or Wightman functions will be in the form of associated Legendre functions \cite{strominger-conformal,Akhmedov}. The locality of the wavefunction-squared can be expressed in terms of normal derivatives of a Green's function evaluated at the far future having the form of Dirac delta and derivatives. Similarly, the particle production is related to the coincident limit of an $in-out$ Green's function. As we will point out in Section \ref{sec:general}, the relationship between the two is not evident, and our main discussion will not utilize them.

Instead we will focus on the wavefunction and particle production of heavy fields on the global and Poincare patch de Sitter. While the local wavefunction-squared result was produced originally for the massless case, the particle interpretation of massless fields is non-trivial at the far future of de Sitter as they do not have an oscillatory behaviour.\footnote{Or more precisely the phase of the oscillation becomes large.} Because of this we will focus on the heavy masses where the particle interpretation is much easier, and indeed will see that there will be a case where their wavefunction-squared is local.

We begin in the next section by defining the concepts of wavefunction and the particle production due to a curved background geometry. We will show that the locality result of \cite{maldacena-non-gaussian} on the Poincare patch can be generalized to other light masses. We will recall the ambiguity of the choice of vacuum on curved spacetimes, the choice of $in$ and $out$ vacuum for the heavy masses and the concept of particle creation. Next in Section \ref{sec:global} we will study the free heavy masses on the global de Sitter and show that they have local wavefunction-squared in odd dimensions at far future if one starts off from the $in$ vacuum-proper for the global de Sitter. We will see that this behaviour is directly connected to the absence of particle production noted in \cite{strominger-conformal}, by writing the $in$ mode in terms of the $out$ mode and its conjugate. 

Next, in Section \ref{sec:poincare} we will discuss the Poincare patch. Note that this situation is different from that of global de Sitter, mainly because the $in$ modes are different. Here we will see that the wavefunction is always non-local and there is always particle production for the heavy masses. We will again note that both phenomena arise from the relationship of the $in$ mode to the $out$ mode. In Section \ref{sec:general} we will point out that the correspondence between the absence of particle production and locality of the wavefunction-squared is not special to de Sitter. More specifically we show that for systems in which the $out$ mode has a separated time and momentum dependency at far future, no particle production implies local wavefunction-squared at far future. We discuss for which geometries this happens, and see that this will hold for most expanding geometries. We note that the reverse direction of the correspondence is harder to prove, but expected to hold in most circumstances. Additionally we comment on how both the particle production and wavefunction-squared can be expressed in terms of Green's functions and discuss the difference between these two Green's functions. We conclude with Section \ref{sec:Conc}, where we discuss further extensions of the result and comment on some related results in the literature.

\section{The Wavefunction, Locality and Particle Production}\label{sec:Basics}

Let us state the problem here more precisely. In a quantum field theory, the wavefunction is the probability of a certain field configuration to be realized in a given state of the system at a given time. More specifically, for a system that is in the state $\left| \alpha \bra$ at a time $t$, it is simply
\begin{equation}
\Psi_\alpha[\varphi(\vec{x})] := \ket \Phi(t,\vec{x}) | \alpha \bra
\end{equation}
where $\left| \Phi(t,\vec{x}) \right>$ is the eigenstate of the canonical field $\Phi(t,\vec{x})$ with eigenvalue $\varphi(\vec{x})$ at time $t$. For cosmological purposes one typically chooses the state to be a vacuum at an initial time; we will focus on such cases. Let us consider free fields on a manifold $\mathcal{M}=[t_i,t_f] \times M$ with a metric
\begin{equation} \label{eq:metric}
ds^2=-a^2(t) \, dt^2+\sigma^2(t) \, \gamma_{ij}(\vx) \, dx^idx^j  \ .
\end{equation}
For this case one can write the norm of the wavefunction, or the wavefunction-squared, as
\begin{equation}\label{eq:WF1}
\left| \Psi_\alpha[\varphi(\vec{x})] \right|^2= \left| \mathcal{N} \right|^2 \exp \lp - \frac{1}{4} \int dV_x \ dV_y \ \varphi(\vec{x}) \, \varphi(\vec{y}) \int d^{d-1}k \ Y_\vk(\vx) Y^*_\vk(\vy) \ \frac{1}{|{\phi_\alpha}(t,k)|^2 }  \ \rp
\end{equation}
where $\mathcal{N}$ is a normalization independent of $\varphi(\vec{x})$, $dV_x$ is the Riemannian volume form for $\gamma$ in coordinates $\vx$, $\phi_\alpha$ is the mode function corresponding to the $\alpha$ vacuum and $\lbrace Y_\vk(\vx) \rbrace$ forms a basis of functions on $M$. See appendices for more explanation on the setup and details. One choice of vacuum, which we will denote with $\alpha=i$, is special; it is the vacuum with the mode functions such that $\phi_i^*(t,k)$ vanishes at the far past $t_i$. For the Poincare patch of de Sitter this is the Bunch-Davies vacuum with the corresponding modes
\begin{equation}
\phi_i(t,k)=\sqrt{\frac{\pi}{4}} e^{-\pi \mu/2} (-H t)^{(d-2)/2} \sqrt{-t} \ H^{(1)}_{\nu}(-k t) \gwg \nu=\sqrt{\frac{(d-1)^2}{4}-\frac{m^2}{H^2}} \ ,
\end{equation}
where $H$ is the Hubble constant. With the correct $i\epsilon$ prescription one can show that this vanishes as $t \ra t_i$ on the Poincare patch. For this choice of vacuum as the state of the system, it has been noted that when the mass is zero and the dimension is odd $|{\phi_i}(t_f,k)|^2\sim k^{-2n}$ for an integer $n$, thus the wavefunction-square is \textit{local} \cite{maldacena-non-gaussian}.\footnote{As the momentum integral inside eq \eqref{eq:WF1} will produce Dirac delta and derivatives of Dirac delta.} This result can be generalized to light masses by noting
\begin{equation}
\phi_i(t,k)\sim H^{(1)}_{\nu}(-k t) \ra (-kt)^{-\nu}
\end{equation}
where locality happens when $\nu$ is an integer.\\
\hfill\\
On a curved spacetime, due to the changing background, even for free fields there will be particle production. Phrased in the canonical quantization language, this is due to the fact that there are different mode functions $\phi_\alpha(t,k)$ and corresponding creation and annihilation operators $\adk$ and $\ak$ with which one can write
\begin{equation}
\Phi(t,\vec{x}) = \int d^{d-1}k \lp \phi_\alpha(t,k) \, \ak \, Y_\vk(\vx) + \phi^*_\alpha(t,k) \, \adk \, Y_\vk^*(\vx) \rp \ .
\end{equation}
A given choice of mode function and creation operator will correspond to different quanta, or a different prescription of a particle. This brings about the well known problem of defining what a particle is on a curved spacetime. The quanta corresponding to the Bunch-Davies vacuum described above for the Poincare patch are considered to be an accurate identification of particles at the far past--as the corresponding modes behave as they would in flat space, namely,
\begin{equation}
\lim_{t \ra t_i} \phi_i(t,k) \sim \frac{1}{\sqrt{k}} \ e^{-ikt} \ .
\end{equation}
Motivated by this, one can impose that the correct description of a particle at the far past and far future correspond to the choice of $\left|in\bra$ and $\left|out\bra$ vacua such that
\begin{equation}
\label{eq:in-beh}
\phi_{in}(t,k) \sim e^{-i \, \omega_{in} \, t} \quad \quad t \ra t_i
\end{equation}
and
\begin{equation}
\label{eq:out-beh}
\phi_{out}(t,k) \sim e^{-i \, \omega_{out} \, t} \quad \quad t \ra t_f \ ,
\end{equation}
where $\omega_{in},\omega_{out}$ are some positive frequencies that can in practice depend on time and momentum. Note that one must also require the mode functions to satisfy the Wronskian \eqref{eq:Wr2} for canonical quantization; this is also the reason for the sign in front of the time in the exponential. These conditions will uniquely fix the $in$ and $out$ modes as the other independent solution in the respective regions can only be the complex conjugate. Note that for the Poincare patch of de Sitter $\phi_{in}=\phi_{i}$.
If the system starts out at $\left| in \right>$ vacuum, there will in general be a non-zero probability to find particles of the $\left| out \right>$ vacuum in it. Assuming these two solutions are complex \footnote{As opposed, for example, to the case where $\phi_{out} \sim J_{\nu}$, which one might attempt to choose in analogy to the heavy mass case that will be discussed in Section \ref{sec:poincare}.}  so that a solution and its complex conjugate are linearly independent, one can always write the $in$ solution in terms of the $out$ solution, namely 
\begin{equation}\label{eq:Bogol1}
\phi_{in}(t,k)= \alpha_k \ \phi_{out}(t,k) + \beta_k \ \phi^*_{out}(t,k)
\end{equation}
for some complex constants $\alpha_k,\beta_k$. Note that the Wronskian condition requires
\begin{equation}
|\alpha_k|^2-|\beta_k|^2=1 \ .
\end{equation}
$\beta_k$ is a measure of how much $in$ vacuum is different from the $out$ vacuum, thus if $\beta_k$ is zero one concludes there is no particle creation at far future. As we will discuss in the next section and as was pointed out in \cite{strominger-conformal}, in global de Sitter there is no particle creation in odd dimensions.

\section{The Case of Global de Sitter}\label{sec:global}

In this section we will consider the case of global de Sitter of arbitrary dimension and show that in odd dimensions the wavefunction is local and there is no particle production. We will observe that both of these facts come about by an identity that connects the $in$ mode functions to the $out$ mode functions.

We will work in the global coordinates $(\tau,x^A)$ such that
\begin{equation}
ds^2=-d\tau^2+\cosh^2(\tau) \ d \Omega^2_{d-1} = -d\tau^2+\cosh^2(\tau) \ \gamma_{AB}(x^C) dx^A dx^B 
\end{equation}
where $\gamma$ is the round metric on $S^{d-1}$, which we will refrain from writing explicitly. Note that $\tau_f=\infty$ and $\tau_i=-\infty$. We will consider the free field with a mass with the action
\begin{equation}
S= \frac{1}{2} \int d\tau \ d^{d-1}x \sqrt{\gamma} \, (\cosh\tau)^{(d-1)} \lp \dot{\phi}^2- \frac{(D\phi)^2}{\cosh^2\tau}-m^2\phi^2 \rp \ ,
\end{equation}
dot denoting the $\tau$ derivative and $D$ being the $\gamma$-covariant derivative. We will expand our quantum fields on the closed constant time slices as
\begin{equation}
\phi(\tau,\Omega)= \sum_{Lj} \lp a_{Lj} \, y_L(\tau) \, Y_{Lj}(\Omega) + a_{Lj}^\dagger \, y^*_L(\tau) \, Y^*_{Lj}(\Omega) \rp
\end{equation}
where $L$ is a non-negative integer and $j$ denotes the rest of the indices collectively. We will follow the conventions of \cite{strominger-conformal} in spherical harmonics $Y_{Lj}(x^A)$. Note that 
\begin{equation}
D^2Y_{Lj}(x^A)=-L(L+d-2)Y_{Lj}(x^A) \ .
\end{equation}
The canonical commutation relations require \eqref{eq:Wr2}, or
\begin{equation}
y_L(\tau)  \dot{y}^*_L(\tau)-y^*_L(\tau) \dot{y}_L(\tau)= \frac{i}{ (\cosh\tau)^{(d-1)}} \ . \label{eq:Wronskian-global}
\end{equation}
The wavefunction-squared for a state chosen as a vacuum denoted by $\alpha$, at time $\tau_f$ for a field configuration $ \varphi(\Omega)$ is given by
\begin{equation}
\left| \Psi_\alpha[\varphi(\Omega)] \right|^2= \exp \lp -\frac{1}{4} \int d\Omega_x \, d\Omega_y \, \varphi(\Omega_x) \varphi(\Omega_y) \, \sum_{Lj} Y_{Lj}(\Omega_x) Y_{Lj}(\Omega_y) \, \frac{1}{\left| y^{\alpha}_{L}(\tau_f) \right|^2} \rp \ .
\end{equation}
Now we discuss the solution $y^{in}_L$. This is the solution to the free field equations that satisfies \eqref{eq:in-beh}, also defined in \cite{strominger-conformal}. One can write this as
\begin{equation}
y^{in}_L(\tau)= \frac{\Gamma(1-i\mu)}{\sqrt{2\mu}} \frac{P^{i\mu}_{\lambda}(-\tanh\tau)}{ \lp \cosh\tau \rp^{\frac{(d-1)}{2}}} \gwg \lambda=L+\frac{(d-3)}{2}, \ \mu=\sqrt{m^2-\frac{(d-1)^2}{4}} 
\end{equation}
and $P_\lambda^{i\mu}$ is the associated Legendre function. With the correct $i\epsilon$ prescription $-\infty \ra -\infty(1+i\epsilon)$, $y^{in}_L(\tau_i)$ goes to zero as $\tau_i \ra -\infty$, so it is also the mode that vanishes at $\tau_i$. One can check that the normalization is chosen such that the Wronskian \eqref{eq:Wronskian-global} is satisfied using identity \eqref{eq:P-Wronskian}. \\
Similarly the out function satisfying \eqref{eq:out-beh} and the correct normalization is written as
\begin{equation}
y^{out}_L(\tau)= \frac{\Gamma(1+i\mu)}{\sqrt{2\mu}} \frac{P^{-i\mu}_{\lambda}(\tanh\tau)}{ \lp \cosh\tau \rp^{\frac{(d-1)}{2}}} \ .
\end{equation}
Note that $y^{out}_L(\tau)={y^{in}}_L^*(-\tau)$.
The crucial identity for us will be 
\begin{equation} \label{eq:Bogol}
y^{in}_L(\tau)= \lp \frac{\sin(\lambda \pi)}{i \tanh(\mu \pi)} + \cos(\lambda \pi) \rp e^{i (\delta_L-\delta_\mu)} \, y^{out}_L(\tau) - \frac{\sin(\lambda \pi)}{i \sinh(\mu \pi)} \, {y^{out}_L}^*(\tau) \ ,
\end{equation}
which can be derived using a reflection identity of Legendre functions \eqref{eq:P-ref}. Here we have defined the phases $\delta_\mu$ and $\delta_L$ such that
\begin{equation}
e^{i\delta_\mu}= \frac{\Gamma(1+i \mu)}{\Gamma(1-i\mu)} \gag e^{i{\delta}_L}= \frac{\Gamma(\lambda+i\mu+1)}{\Gamma(\lambda-i\mu+1)} \ .
\end{equation}
Note that \eqref{eq:Bogol} is essentially $in$ modes written in terms of the $out$ modes with proper Bogoliubov coefficients. As noted in \cite{strominger-conformal}, for odd dimensions where $\lambda=L+(d-3)/2$ is an integer, the $\beta$ coefficient is zero, so there is no particle production at the far future of dS: put in another way, if the system is in the $in$ vacuum state at far past there will be no $out$ particles in the far future. Interestingly enough exactly the same identity will imply that the wavefunction square will be local at far future. At the root of this surprise lies the fact that $y^{out}_L$ is \textit{the} mode that has an $L$-independent amplitude at far future. More specifically as $\tau \ra \infty$, $ \tanh \tau \ra 1-2 e^{-2\tau}$, and using the behaviour of Legendre P around $1^-$, \eqref{eq:1minus-limit}, we see that
\begin{equation}\label{out-Global}
\lim_{\tau \ra \infty} y^{out}_L(\tau) = \frac{e^{i\delta_\mu}}{\sqrt{2\mu} (\cosh\tau)^{\frac{(d-1)}{2}}} \, e^{-i \mu \tau} \ .
\end{equation}
For odd dimensions $y^{in}_L \sim y^{out}_L$ and the amplitude is independent of $L$, namely
\begin{equation}
\lim_{\tau_f \ra \infty} \left| y^{in}_L(\tau_f) \right|^2= \frac{1}{2\mu} \frac{1}{\lp \cosh\tau_f \rp^{(d-1)}}
\end{equation}
producing 
\begin{equation}
\sum_{Lj} Y_{Lj}(\Omega_x) Y_{Lj}(\Omega_y) \, \frac{1}{\left| y^{in}_{L}(\tau_f) \right|^2}= 2\mu (\cosh\tau_f)^{(d-1)} \delta^{(d-1)}(\Omega_x-\Omega_y)
\end{equation}
which in turn implies a local wavefunction-squared. Meanwhile for even dimensions we have the mixing of the two oscillatory modes $e^{\pm i\mu\tau}$ with $L$-dependent coefficients, so that
\begin{equation}
\left| y_L(\tau_f) \right|^2= \frac{\lp 1+ \cosh^2(\pi \mu)-2 \cosh(\pi \mu)\cos(2 \mu \tau_f-\delta_L+\delta_\mu) \rp}{2\mu \lp \cos\tau_f \rp^{(d-1)} \sinh^2(\pi\mu)} \ .
\end{equation}
Since $L$- dependence does not disappear and is not in the form that would produce the derivatives of the Dirac delta, we will have a non-local wavefunction squared for this case.

To summarize, we have seen that the wavefunction-squared is local at $\tau_f$ precisely when there is no particle production. As we have pointed out, this is because $y_L^{out}(\tau)$ is \textit{the} mode that has L, or momentum, independent amplitude at far future. Looking closer, this seems true almost by definition: if a mode has the behaviour $y_L^\alpha(\tau) \sim e^{-i \omega_\alpha \tau}$ at a time $\tau$, where $\omega_\alpha$ is a constant that is independent of both $L$ and $\tau$, then its amplitude has to be momentum independent due to the Wronskian condition. Note that for a field that satisfies the canonical commutation relations, the Wronskian itself will always be momentum-independent. We will explore the scope of this observation in Section \ref{sec:general}. Now we turn to the Poincare patch.

\section{The Poincare Patch}\label{sec:poincare}
The result of no particle creation in odd dimensional space was claimed to be reproduced for the Poincare patch in \cite{Akhmedov}. However one can easily show that the wavefunction for heavy masses is non-local for any dimensions and mass, implying the breakdown of the correspondence observed in the previous section. Nevertheless the simple calculation of the $\beta$ coefficient affirms our claim, showing that there is always particle production in the Bogoliubov sense. We comment on the discrepancy between our result and the result of \cite{Akhmedov} in Section \ref{sec:general}.

The Poincare patch is best described in conformal coordinates where
\begin{equation}
ds^2=(H\eta)^{-2} (-d\eta^2+d\vec{x}^2) \ .
\end{equation}
In this case $\eta_f=0$ and $\eta_i=-\infty$. The free field action then takes the form 
\begin{equation}\label{eq:free-action-Poincare}
S=\frac{1}{2}\int d\eta \ d^{(d-1)}x \ (H\eta)^{-(d-2)} \left(  \Phi'^2-(\partial\Phi)^2- \frac{m^2}{(H\eta)^2} \Phi^2 \right) \ .
\end{equation}
The constant time slices are open and flat, $Y_\vk(\vx)=e^{i \vk.\vx}/\sqrt{2\pi}$ and the field is quantized as usual by writing
\begin{equation}
\phi(\eta,\vx)= \int \frac{d^{d-1}k}{\sqrt{2\pi}} \lp a_\vk \, \phi_{in}(\eta,k) \, e^{i \vk.\vx} + a_{\vk}^\dagger \,  \phi^*_{in}(\eta,k) \, e^{-i \vk.\vx} \rp \ .
\end{equation}
The Wronskian equation becomes
\begin{equation}
\phi_{in}(\eta,k)  \dot{\phi}_{in}^*(\eta,k)-\phi_{in}^*(\eta,k) \dot{\phi}_{in}(\eta,k)= i (-H\eta)^{d-1} \ . \label{eq:Wronskian-Poincare}
\end{equation}
The wavefunction-squared at time $\eta_f$ for a field configuration $\varphi(\vx)$ is given by
\begin{equation}
\left| \Psi_\alpha[\varphi(\vec{x})] \right|^2= \left| \mathcal{N} \right|^2 \exp \lp - \frac{1}{4} \int dV_x \ dV_y \ \varphi(\vec{x}) \, \varphi(\vec{y}) \int \frac{d^{d-1}k}{2\pi} \ e^{i\vk \cdot (\vx-\vy)} \ \frac{1}{|{\phi_{in}}(\eta, k)|^2}  \ \rp \ .
\end{equation}
We will consider again the case where the mass is heavy and we define
\begin{equation}
\mu=\sqrt{\frac{m^2}{H^2}-\frac{(d-1)^2}{4}} \ .
\end{equation}
Then the $in$ modes are written as
\begin{equation}
\phi_{in}(\eta,k)=\sqrt{\frac{\pi}{4}} e^{-\pi \mu/2} (-H\eta)^{(d-2)/2} \sqrt{-\eta} \ H^{(1)}_{i\mu}(-k\eta) 
\end{equation}
whereas
\begin{equation}
\phi_{out}(\eta,k)=\sqrt{\frac{\pi}{2\sinh(\pi \mu)}}  (-H\eta)^{(d-2)/2} \sqrt{-\eta} \ J_{i\mu}(-k\eta) \ .
\end{equation}
This is because the Bessel functions behave as
\begin{equation}
\lim_{\eta \ra \eta_i} H^{(1)*}_{i\mu}(-k\eta) \ra \sqrt{\frac{2}{-\pi k \eta}} e^{i k \eta_i(1-i\epsilon)} \ra 0
\end{equation}
and
\begin{equation}\label{eq:out-Poincare}
\lim_{\eta \ra \eta_f} J_{i\mu}(-k\eta) \ra \lp \frac{-k \eta_f}{2} \rp^{i\mu} \frac{1}{\Gamma(i\mu+1)} \sim e^{i\mu \ln(-\eta)} \sim e^{-i\mu t}
\end{equation}
where $t$ is the cosmological time. Note that in this case the conformal time dependency of the factor of the exponential is more complicated than a simple linear dependency. Nevertheless we see that the amplitude of the $out$ mode is momentum independent at far future.

The relationship between the $in$ and $out$ modes is simply obtained from the definitions of the Hankel and Bessel functions and reads as
\begin{equation}
\phi_{in}(\eta,k)= \frac{e^{\pi \mu/2}}{\sqrt{2 \sinh(\pi\mu)}} \phi_{out}(\eta,k)-\frac{e^{-\pi \mu/2}}{\sqrt{2 \sinh(\pi\mu)}} \phi_{out}^*(\eta,k) \ .
\end{equation}
Note that unlike the global case there is no finite value of $\mu$ for which the $\beta$ coefficient is zero, thus on the Poincare patch there will always be production of heavy masses. Consequently at the far future $\phi_{in}(\eta,k)$ will always be a mixture of two oscillators and the wavefunction will be non-local, namely
\begin{equation}
Re[i S_{cl}]\sim \frac{1}{(-H\eta)^{(d-2)}} \frac{1}{\eta} \frac{\mu \sinh(\pi \mu)}{ \cos \lp 2 \mu \ln \lp \frac{-k\eta}{2}\rp -2 \delta_\mu \rp + \cosh(\pi \mu)} \ .
\end{equation}
Note that interestingly global de Sitter behaviour is similar to Poincare patch behaviour only in even dimensions. This suggests there is something drastically different happening in odd dimensional global de Sitter. 

\section{Generalization and Relation to the Green's functions}\label{sec:general}
\paragraph{Generalization to other geometries.}
Seeing in the last section that a time dependence of the $out$ mode at far future more general than that of $e^{-i \omega t}$ still allows for momentum independent amplitude, we check how general a statement can be made in this section. Note that we focus on the condition on the phase of the $out$ mode at far future as this was the defining property of the $out$ mode, and the momentum independence arose as a consequence of the Wronskian condition. We will also go beyond de Sitter and consider the metric \eqref{eq:metric}. We will have in mind the heavy masses such that the solutions have an oscillatory behaviour at the far future and far past. Consideration of other solutions (e.g. light masses) is beyond the scope of this work. 

Let us now consider what the most general $out$ mode would be such that it has momentum independent amplitude at far future. One can always write it in the form
\begin{equation}
\phi_{out}(t,k) = \A(t,k) \, e^{-i \theta(t,k)} 
\end{equation}
where $\A$ and $\theta$ are real. The Wronskian condition implies
\begin{equation}
\A^2(t,k) \, \dot{\theta}(t,k)=i W(t)  
\end{equation}
for some function $W(t)$ independent of $k$; if the field and the conjugate momenta are to satisfy the canonical commutation relations. Thus we see that when $\dot{\theta}(t,k)$ is $k$-independent, or when $\theta(t,k)$ has ``separated'' $k$ and $t$ dependence, the amplitude of the out mode will be $k$ independent. Note that, for example, this will not be true for the flat space or Bunch-Davies vacuum at the far past. The most general statement is:\vspace{3mm} \\
\textit{For systems that have an out mode such that $\partial_t Arg(\phi_{out})$ is momentum independent at the far future, no particle production implies the locality of wavefunction-squared at far future for the free theory.} \vspace{3mm} \\
Indeed, writing the amplitude-squared of the $in$ modes at $t_f$ in terms of the $out$ modes,
\begin{equation}
|\phi_{in}(k,t_f)|^2 = \A^2 (1+ 2 |\beta_k|^2 + \alpha_k \beta_k^* e^{2i \theta} + \beta_k \alpha_k^* e^{-2i\theta} ) \ ,
\end{equation}
we see that $|\phi_{in}(t_f,k)|^2$ is k independent when $\beta_k=0$ and $\A$ is independent of k, producing a local wavefunction. 
The de Sitter space has this property mostly because at late times space dependence becomes suppressed with respect to the acceleration of the universe. However this is true for other geometries too; in terms of the components of the metric \eqref{eq:metric}, choosing coordinates such that $a(t)=\sigma(t)$, what we need is \footnote{Note that in our notation even when we choose conformal time $\dot{a}$ corresponds to derivative with respect to it, not e.g. the cosmological time.}
\begin{equation}\label{eq:gg}
m^2a^2 - \frac{(d-2)}{2} \frac{\ddot{a}}{a} -\frac{d(d-2)}{4} \frac{\dot{a}^2}{a^2} \gg \lambda_k \ ,
\end{equation}
where $-\lambda_k$ are eigenvalues of $D^2=\gamma^{ij} D_i D_j$, where $D$ is the $\gamma$-covariant derivative. For de Sitter, $\frac{\ddot{a}}{a},\frac{\dot{a}^2}{a^2} \sim a^2$ and the inequality will hold at far future as $a \ra \infty$. For an FRW power-law cosmology where $a \sim \eta^p$ and $\eta_f \ra \infty$ we will have $\frac{\ddot{a}}{a},\frac{\dot{a}^2}{a^2} \ra 0$ and mass term will satisfy the inequality by itself.

We conclude by noting that above we have found the condition on the $out$ mode under which the absence of particle production imply the locality of the wavefunction-squared. The reverse implication is also expected to hold, but harder to prove: momentum independency of $|\phi_{in}(k,t_f)|^2$ might imply that $\beta_k$ is momentum independent, not necessarily that it vanishes.

\paragraph{Connection to the Green's functions.}

Now we will connect our discussion to the two commonly discussed objects in the literature; the in-out propagator discussed in QFT on dS literature \cite{Akhmedov} and the bulk-to-bulk propagator discussed in the holography \cite{Witten98, Giddings99} and the cosmological correlator literature \cite{Pajer-optical}.

The bulk-to-bulk propagator $G_{BB}$ is defined as the Green's function that vanishes on the boundary of the manifold, see Appendix \ref{app-B} for more details. In the space of basis functions it can be written in terms of the mode functions as
\begin{equation}
\gbb(k,t,t')=- \lp \ \theta(t-t') \phi_i^*(t',k) \phi_f(t,k) + \theta(t'-t) \phi_i^*(t,k) \phi_f(t',k) \ \rp
\end{equation}
where $\phi^*_i(t',k)$ is what we will call the $initial$ mode function, one that vanishes at $t_i$, and $\phi_f(t,k)$ is the $final$ mode function, one that vanishes at $t_f$. It is related to the wavefunction via \eqref{eq:WF} and \eqref{eq:B-Bnd to BB}, or as
\begin{equation}\label{eq:WF-gbb}
\left| \Psi_{i}[ \varphi(\vx) ]\right|^2 =|\mathcal{N}|^2 \exp\lp -\frac{1}{2}\int dV_x dV_y \ \varphi(\vx) \varphi(\vy) \ \frac{\sigma^{2(d-1)}(t_f)}{ a^{2}(t_f) } \Re\lp \left. \pr_t \pr_{t'} G_{BB}(t,\vx,t',\vy) \right|_{\substack{t=t_f\\t'=tf}} \rp\rp \ .
\end{equation}
Thus wavefunction-squared will be local if the $\Re(\pr_t \pr_t G_{BB})$ is Dirac delta or derivatives.

One can also connect the particle creation to a Green's function. This is known for Schwinger particle production and in the cosmological context was discussed in \cite{Akhmedov}. The Green's function relevant here is simply the $in-out$ ``propagator'' $G_{in-out}$. It is defined as
\begin{equation}
G_{in-out}(t,\vx,t',\vx'):= \frac{\ket out \right| T \, \Phi(t,\vx) \Phi(t',\vx') \left| in \bra}{\ket out|in \bra}
\end{equation}
where $T$ is the time ordering symbol. In the space of basis functions $\gio$ is written as
\begin{equation} \label{eq:gio}
\gio(t,t',k)= \frac{1}{\alpha_k^*} \lp \theta(t-t') \phi^*_{in}(t',k) \phi_{out}(t,k) +  \theta(t'-t) \phi^*_{in}(t,k) \phi_{out}(t',k)  \rp
\end{equation}
where $\alpha_k$ is defined in \eqref{eq:Bogol1}. The particle creation is linked to the imaginary part of the coincidence limit of $\gio$ in the position space. Easiest way to see this is through writing the $out$ modes in terms of the $in$ modes inside $\gio$ to get
\begin{equation}\label{eq:gio-beta}
G_{in-out}(t,\vx;t',\vx')=G_{in-in}(t,\vx;t',\vx')+\int d^{(d-1)}k \frac{\beta_k}{\alpha^*_k} \ Y_k(\vx) Y_k^*(\vx') \phi_{in}^*(t,k) \phi_{in}^*(t',k)
\end{equation}
Then one notices that the imaginary part of $G_{in-in}$ is zero in the coincident limit; thus \linebreak $\Im(\gio(t,\vx,t,\vx))=0$ implies $\beta_k=0$ and vice versa.\footnote{One might argue this is true only if $\phi^*_{in}(t,k)$ is not zero at $t$. However $G_{in-out}$ is in fact a function of the invariant distance, which is equal to 1 at all coincident limits; and thus its value should not depend on a specific time instance.} In \cite{Akhmedov}, it was argued that $\Im(\gio(t,\vx,t,\vx))=0$ in odd dimensions on the Poincare patch for heavy masses. However in Section \ref{sec:poincare} we have seen that $\beta_k$ is never zero. The discrepancy is caused by the fact that they regularize $G_{in-out}$, but this doesn't affect our results as long as we understand the particle creation in the Bogoliubov sense. We comment on this further in the next section. 

Before we conclude, note that the particle creation is related to $in$ and $out$ mode functions rather than the $initial$ and $final$ mode functions. Using a proper $i\epsilon$ prescription renders the $in$ and $initial$ mode functions equal to each other as we have discussed, while the same does not seem necessarily to be true for the $final$ and $out$ mode functions. 

One might object to this, pointing out that the both of the $out$ modes of global dS and Poincare patch dS, \eqref{out-Global} and \eqref{eq:out-Poincare}, vanish as $t \ra t_f$. However the definition \eqref{eq:final-mode-def},  ensuring $\phi_f$ vanishes at a time $t^-_f<t_f$, before one takes the limit $t^-_f \ra t_f$, excludes the possibility of $\phi_f=\phi_{out}$. This is because while $\phi_{out}$ needs to satisfy the Wronskian condition \eqref{eq:Wr2}, $\phi_f$ cannot by definition, as the Wronskian is zero for this case.

\section{Conclusions and Discussion}\label{sec:Conc}

We have shown that both in global dS and the Poincare patch dS, no particle production implies locality of the wavefunction-squared at far future for free heavy fields. We have seen that this is realized in the odd dimensional global de Sitter but never in the Poincare patch. Furthermore we have shown that the correspondence is not a peculiarity of de Sitter; it holds whenever the $out$ modes have a phase with separated dependence in time and momentum. We have pointed out that this would be the case, for example, for geometries expanding with a power law. We have noted that even though it is easy to show how no particle production implies locality of the wavefunction-squared, the reverse implication is not trivial to prove. Now let us discuss some interesting directions worth further inquiry.

\paragraph{``Transparency'' of odd dimensional global de Sitter.}The global de Sitter of odd dimensions is the only place where we have seen that there is no particle production and thus the wavefunction-squared is local, for the heavy masses. A similar result does not occur in the Poincare patch, and the main reason for this seems to be that the respective $in$ modes are different. Nevertheless odd-dimensional global de Sitter looks rather special when the system starts off in the $in$ vacuum state. It is an interesting question why this happens, some work that addresses this issue are \cite{morrison,maloney,behrang}, where the phenomenon was called the transparency of odd dimensional de Sitter. 

\paragraph{Light masses.} Note that when discussing locality, we have only focused on the case where a Dirac delta multiplies $\varphi(\vx) \, \varphi(\vy)$ in the action, not derivatives of Dirac delta, which actually still produces a local action. This is mostly because there doesn't seem to be a natural way for this wto happen for a heavy mass. If one is to consider light masses, this is indeed how locality appears as we have pointed out in Section \ref{sec:Basics}. However in this case there is no well-known way to describe what a particle is at the far future, thus one cannot describe a correspondence. The main reason is that for the light masses the oscillatory behaviour does not exist anymore.

The analogy with the harmonic oscillator can still be made as long as there is a minimum of the energy at the far future \cite{MW}. For dS this will be the case only for masses heavier than the conformal mass, i.e., when $\nu \le 1/2$. On the Poincare patch one has $|{\phi_i}(t_f,k)|^2\sim k^{-2 \nu}$, which is never local for these values of $\nu$. For the conformal mass $\nu=1/2$, the Bunch-Davies choice will still minimize the energy, thus when taken as the $out$ mode there will be no particle production, breaking the correspondence that existed for the heavy mass. Note that this reflects exactly the situation in the flat case. 

\paragraph{Green's functions.} In the last section we have shown that $\Im\lp G_{in-out}(x,t;x,t) \rp \neq 0$ on Poincare patch due to the existence of particle creation, i.e. $\beta_k \neq 0$ for all dimensions and pointed out that this contradicts the result of \cite{Akhmedov}, mainly because they do not calculate the bare $G_{in-out}$, but the regularized one. The reason for this seems to be the fact that the bare $G_{in-out}$ diverges in the coincident limit. From their discussion it is not clear whether the imaginary part is also divergent, as this is generally not expected to be the case for example, in analogous Schwinger effect. The gist of the problem is whether the regularized $G_{in-out}$ or the $\beta_k$ coefficient is the correct measure of particle creation. 

In Section \ref{sec:general} we have discussed the relationship of the locality of the wavefunction-squared to the Green's function $G_{BB}$. We have noted that $G_{BB}$ and $G_{in-out}$ would be the same quantity (possibly up to some overall constant) if $\phi_{out}=\phi_f$. The only reason that this is not the case is the insistence on wanting $\phi_f$ to vanish at a time $t^-_f<t_f$ before taking the limit $t^-_f \ra t_f$. Note that this is not necessary for the formula of eq. \eqref{eq:WF-gbb} to be correct. As it can be seen by following the discussion of Appendix \ref{app-B} and writing the covariantized Green's functions, the actual requirement is that a proper combination of volume factors (which would include $a(t)$ and $\sigma(t)$), $\gbb$ and $\partial_t K$ vanish as $t \ra t_f$ (the vanishing boundary condition is automatically satisfied at $t_i$ with the correct $i\epsilon$ prescription.). In fact the condition
\begin{equation}
\lim_{t \ra t_f} G_{BB} = 0
\end{equation}
is not sufficient unless $t_f$ is replaced by $t^-_f$. Note however that $t \ra t_f$ limit is non-trivial for de Sitter because of the infinitely fast oscillating factors and presumably a regularization is needed. We leave these concerns for future study.

\acknowledgments

The problem was suggested to me by Guilherme L. Pimentel; I would like to thank him for many discussions and supervision. I would like to thank Craig R. Clark and Tom Westerdijk for the initial collaboration and many discussions; Sumer Jaitly, Behrang Tafreshi, Carlos Duaso Pueyo, Facundo Rost, Ugo Moschella, Ali Seraj and Silvia Pla for discussions. Finally I would like to thank Guilherme L. Pimentel, Craig R. Clark, Tom Westerdijk and Behrang Tafreshi for a review of the manuscript. I am financially supported by Scuola Normale Superiore, INFN (IS-GSS) and ERC (NOTIMEFORCOSMO, 101126304) grant funded by the European Union. 

\appendix

\section{Quantum Fields on Curved Spacetimes, Vacuum Choice, Particle Production and the Wavefunction}
\label{sec:qft-CS}
In this appendix we review some features of the quantum fields on de Sitter space and set up our conventions.\footnote{For more details see \cite{baumann, MW, Weinberg, Kutluk_notes}.} We will mostly have the observables of quantum cosmological perturbations in the very early universe in mind. Assume that the universe has started out in a state $\left|\psi \bra $ in the far past $t_i$ and we are interested in the observable $\obs$ at time $t$. This would be measured as the expectation value
\begin{equation}
\ket \obs \bra(t) := \ket \psi | \obs(t) | \psi \bra \ .
\end{equation}
In the Heisenberg picture the state $\left| \psi \bra$ is constant and operators evolves with the evolution operator $U(t,t_i)$ defined as the unique operator that solves
\begin{equation}
\frac{d U(t,t_i)}{dt}=-i H(t)U(t,t_i)
\end{equation} 
with the boundary condition $U(t_i,t_i)=1$, so one has \cite{Kutluk_notes}
\begin{equation}
\ket \obs(t) \bra = \ket \psi | {U}^{-1}(t,t_i) \obs(t_i) {U}(t,t_i) | \psi \bra \ .
\end{equation}
It is often useful to connect this quantity to the wavefunction $\Psi[\phi(t, \vx)]:=\ket \psi(t) | \phi(t,\vx) \bra$, where $\left|\phi(t,\vx)\bra$ is the eigenvector of the canonical field $\Phi(\vx,t)$, with the eigenvalue $\varphi(\vx)=\phi(t,\vx)$.  This could be done by introducing the identity operator expanded in the field basis at time $t$ so that for $\obs= \Phi(t,\vx_1) \cdots \Phi(t,\vx_2)$
\begin{equation}
\ket \obs(t) \bra = \int D\phi\ \phi(t,\vx_1) \cdots \phi(t,\vx_2) \left| \Psi[\varphi(\vx)] \right|^2
\end{equation}
where
\begin{align}
\Psi[\varphi(\vx)]&=  \ket \psi \right|U^{-1}(t,t_i) \left|\phi(t_i,\vx)\bra \ .
\end{align}
Note that
\begin{equation}
\left| \phi(t_i,\vx) \bra:= U(t,t_i) \left| \phi(t,\vx) \bra
\end{equation}
is the eigenvector of $\Phi(t,\vx)$ at time $t_i$ with eigenvalue $\varphi(\vx)$. Inserting an identity operator in the field basis at time $t_i$ we see
\begin{align}
\Psi[\varphi(\vx)]&=  \int D\tilde{\phi}_{t_i} \ket \psi \right| \left. \tilde{\phi}(t_i,\vx) \bra \ket \tilde{\phi}(t_i,\vx) \right|U(t,t_i) \left| \phi(t,\vx) \bra \\
&= \int D\tilde{\phi}_{t_i} \ket \psi \right| \left. \tilde{\phi}(t_i,\vx) \bra \int \mathcal{D}\phi \, e^{i \left. S \right|^\phi_{\tilde{\phi}}} \ .
\end{align}
\paragraph{Choice of vacuum and particle production.} The state of the system is generically thought to be a vacuum of the free Hamiltonian, in analogy to a set of harmonic oscillators. For a background spacetime with metric $g$ in coordinates $x^\mu= (t,\vx)$ a free action is written as
\begin{equation}
S=\int d^dx \sqrt{-g} \lp -\frac{1}{2} \nabla_\mu \Phi \nabla^\mu \Phi +\frac{1}{2}  m^2 \Phi^2 \rp \ .
\end{equation}
We will focus on the metrics of the form
\begin{equation}
ds^2=-a^2(t)dt^2+\sigma^2(t)\gamma_{ij}(\vx)dx^idx^j  
\end{equation}
in some coordinates $(t,\vx)$. Note that for any Lorentzian metric there exist coordinates such that $g_{tt}=-1$ and $g_{ti}=0$ but assuming $g_{ij}(t,\vx)= \sigma^2(t) \gamma_{ij}(\vx)$ is more restrictive. It implies that the spatial slices are conformally mapped to each other with a constant conformal factor, or more concretely Ricci scalars have the relationship $R_{t_1}(\vx_1)= f(t_1,t_2)R_{t_2}(\vx_2)$. We make this assumption to be able to quantize the field as
\begin{equation}
\Phi(t,\vx) = \int d^{d-1}k \lp \phi(t,k) \, a_{\vk} \, Y_{\vk}(\vx) + \phi^*(t,k) \, a^\dagger_{\vk} \, Y^*_{\vk}(\vx) \rp 
\end{equation}
where the same basis vectors $Y_{\vk}(\vx)$ can be used on every spatial slice. Here $\ak,\adk$ are annihilation and creation operators respectively such that
\begin{equation}
[a_\vk,a^\dagger_{\vk'}]=\delta^{(d-1)}(\vk-\vk')
\end{equation} 
and $Y_{\vk}(\vx)$ are a set of functions that forms a basis on $M$, chosen to be the eigenfunctions of $D^2=\gamma^{ij} D_i D_j$, where $D$ is the $\gamma$-covariant derivative. We assume they are chosen to satisfy
\begin{align}
& \int d^{d-1}k \, Y_{\vk}(\vx) Y^*_{\vk}(\vy)= \frac{\delta^{(d-1)}(\vx-\vy)}{\sqrt{\gamma}(\vx)} \ ,\\
&\int d^{d-1}x \, Y_{\vk}(\vx) Y^*_{\vk'}(\vx)=\delta^{(d-1)}(\vk-\vk') \ ,\\
&\int d^{d-1}k \, Y_{\vk}(\vx) Y^*_{\vk}(\vy) f_k=\int d^{d-1}k \, Y_{\vk}^*(\vx) Y_{\vk}(\vy) f_k \ .
\end{align}
The evolution equation for the operators in the interaction picture is by the free Hamiltonian, thus the mode functions $\phi(t,k)$ should satisfy the free equation of motion. The canonical momentum can be expressed in the same basis so that
\begin{equation}
\frac{\Pi(t,\vx)}{\sqrt{\gamma(\vx)}} = \int d^{d-1}k \lp \pi(t,k) \, a_{\vk} \, Y_{\vk}(\vx) + \pi^*(t,k) \, a^\dagger_{\vk} \, Y^*_{\vk}(\vx) \rp \ .
\end{equation}
The canonical commutation relationships then enforce a normalization of the mode functions $\phi(t,k)$ such that
\begin{equation}
\mathcal{W}[\phi(t,k)]=\phi(t,k) \pi^*(t,k)-\phi^*(t,k) \pi(t,k)= i 
\end{equation} 
or in terms of the derivatives of the mode functions
\begin{equation}\label{eq:Wr2}
\mathcal{W}[\phi(t,k)]=\phi(t,k) \dot{\phi}^*(t,k)-\phi^*(t,k) \dot{\phi}(t,k)= i \, a(t) \sigma^{-(d-1)}(t) \ . 
\end{equation} 
Any complex solution to the equations of motion can be made to satisfy this. More specifically if $\phi_\alpha(t,k)$ is a solution with the correct normalization, then another solution
\begin{equation}
\phi_\beta(t,k) = \alpha_k \, \phi_\alpha(t,k) + \beta_k \, \phi_\alpha^*(t,k)
\end{equation}
satisfies the normalization if $|\alpha_k|^2-|\beta_k|^2=1$. Note the new solution defines different creation and annihilation operators related to the older ones by
\begin{equation}
b_k = \alpha_k \, \ak + \beta_k \, \adk \ .
\end{equation}
Because of this the ``$\alpha$-vacuum'' $\left|\alpha \bra$ defined such that $\ak \left|\alpha \bra=0$ is different from the ``$\beta$-vacuum'' $\left|\beta \bra$ defined such that $\bk \left|\beta \bra =0$, more specifically $\left|\alpha \bra$ will contain $\beta$ ``particles'' and vice versa. Note that underlying all of this is our insistence of considering a particle as a harmonic oscillator. \\
In the flat space the usual mode functions are favourable because they minimize the energy and diagonalize the Hamiltonian. On a curved spacetime this could only be done at a single time instance, i.e. once the state of the system chosen to be an energy minimizing vacuum at one time instance it typically will not be energy minimizing at another time. One exception is odd dimensional global de Sitter as we discuss in the main text.\\
\hfill\\
It is only natural to choose the mode functions such that position or other operators act simply on the state of the system. For cosmological purposes most common and intuitive choice for the state is the free vacuum that minimizes the energy at far past. With the $i\epsilon$ prescription, this corresponds to the mode whose complex conjugate vanishes at the initial time which we will call $\phi_i$. The corresponding wavefunction is simply the path integral
\begin{equation}
\Psi_{i}[\varphi(x)]=\mathcal{N} \int^\varphi_{\tilde{\phi}\ra 0} \mathcal{D}\phi \, e^{i S } \ .
\end{equation}
Note that the $\phi_i$ mode corresponds to the Bunch-Davies vacuum on the Poincare patch, but not on the global de Sitter.
When the action is free and there are no interactions, the path integral can be performed exactly and
\begin{equation}\label{eq:Psi-Scl}
\Psi_{i}[\varphi(\vx)]= \mathcal{N} e^{i S[\phi_{cl}(\vx,t)] }
\end{equation}
where $\phi_{cl}(\vx,t)$ is the solution to the free equations of motion with the boundary conditions
\begin{equation}\label{eq:bc}
\phi_{cl}(\vx,t_i)=0 \gag \phi_{cl}(\vx,t_f)=\varphi(\vx) \ .
\end{equation}
Note that this is only true for a $\phi_i$ whose complex conjugate is exponentially suppressed at the far past, and not for other vacua.

\section{The Free action, Bulk to Bulk and Bulk to Boundary Propagators}\label{app-B}

The classical solution that satisfies the boundary conditions of \eqref{eq:bc} can be written as
\begin{equation}
\phi_{cl}(t,\vx)= \int d^{d-1}y \ \sqrt{g_{\Sigma_t}(t,\vy)} \, \varphi(\vy) \, K(t,\vx,\vy)
\end{equation}
where $g_{\Sigma_t}$ is the induced metric on the spatial slice at time $t$, namely $\sqrt{g_{\Sigma_t}}=\sigma^{(d-1)}(t) \sqrt{\gamma}$ and $K(t,\vx,\vy)$ is the bulk to boundary ``propagator''; solution to the equation of motion such that
\begin{equation}
K(t_i,\vx,\vy)=0 \gag K(t_f,\vx,\vy)= \frac{\delta^{(d-1)}(\vec{x}-\vec{y})}{\sqrt{g_{\Sigma_t}(t_f,\vy)}} \ .
\end{equation}
In terms of the $in$ mode functions this is simply 
\begin{equation}\label{eq:K-mom}
K(t,\vx,\vy)=\int d^{d-1}k \ Y_\vk(\vec{x})Y_\vk^*(\vec{y}) \, \frac{1}{\sigma^{(d-1)}(t_f)} \frac{\phi^*_{i}(t,k)}{\phi^*_{i}(t_f,k)}   \ .
\end{equation}
Using this in the action one gets
\begin{equation}\label{eq:WF}
\Psi_{i}[\varphi(x)]= \mathcal{N} \exp\lp \frac{i}{2} \int d^{d-1}y \, d^{d-1}x \sqrt{g_{\Sigma_t}(t_f,\vy)} \sqrt{g_{\Sigma_t}(t_f,\vx)} \ \varphi(\vx) \varphi(\vy) \left. \lp n^\mu(t,\vx)   \nabla_\mu K(t,\vx,\vy) \rp \right|_{t=t_f} \rp
\end{equation}
where $n$ is the outward pointing normal vector field with length $-1$ on the surface $t=t_f$. The bulk to boundary propagator is naturally connected to the ``Bulk-to-bulk'' propagator $ \gbb $, the Green's function normalized such that
\begin{equation}
\lp \nabla_x^2-m^2 \rp \gbb(x,x')=i \frac{\delta^{(d-1)}(x-x')}{\sqrt{g}}
\end{equation}
with the boundary conditions
\begin{equation}
 \left. \gbb(x,x') \right|_{x\in \partial \mathcal{M}}=\left. \gbb(x,x') \right|_{x'\in \partial \mathcal{M}}=0 \ .
\end{equation}
Note that for the spacetimes with closed spatial slices this reduces to conditions on the initial and final slices, whereas for open slices one needs vanishing boundary conditions also at the spatial infinity. The relationship between the propagators can be seen using Green's theorem on a curved spacetime which produces
\begin{equation}\label{eq:B-Bnd to BB}
i K(t,\vx,\vy)= \left. \lp n^\mu(\vx,t')   \nabla_\mu \gbb(t',\vx,t,\vy) \rp \right|_{t'=t_f} \ .
\end{equation}
Note that using this in \eqref{eq:WF} produces the formula $(30)$ in \cite{Polyakov-07} by Polyakov.\\
In the space of basis functions $\gbb$ can be written as 
\begin{equation}
\gbb(t,t',k)=- \lp \ \theta(t-t') \phi_i^*(t',k) \phi_f(t,k) + \theta(t'-t) \phi_i^*(t,k) \phi_f(t',k) \ \rp
\end{equation}
where
\begin{equation}\label{eq:final-mode-def}
\phi_f(t,k)=\phi_i(t,k)-\frac{\phi_i(t_f,k)}{\phi_i^*(t_f,k)}\phi^*_i(t,k) \ .
\end{equation}
The $\phi_f(t,k)$ is the unique solution to the equations of motion that vanishes at $t_f$ and has a zero Wronskian. This definition is formally implemented by replacing $t_f$ with $t^-_f < t_f$, and taking the limit $t^-_f \ra t_f$. \citep{Pajer-optical}

Getting back to \eqref{eq:WF} and plugging in $\sqrt{g_{\Sigma_t}}=\sigma^{(d-1)}(t) \sqrt{\gamma}$, \eqref{eq:K-mom} for $K$ and taking the modulus we see
\begin{equation}\label{eq:WF-square}
\left| \Psi_{i}[\varphi(\vec{x})] \right|^2= \left| \mathcal{N} \right|^2 \exp \lp - \frac{1}{4} \int dV_x \ dV_y \ \varphi(\vec{x}) \, \varphi(\vec{y}) \int d^{d-1}k \ Y_\vk(\vx) Y^*_\vk(\vy) \ \frac{1}{|{\phi_i}(t,k)|^2 }  \ \rp
\end{equation}
where $dV_x$ is the Riemannian volume form for $\gamma$ in coordinates $\vx$. Here we would like to note that, even though the formula \eqref{eq:Psi-Scl} is only correct for the $in$ vacuum, \eqref{eq:WF-square} is true for other vacua too when $\phi_i$ is replaced by the corresponding modes. This can be seen by noting
\begin{equation}
\ket \phi(t,\vx)| \ak |\alpha \bra =0
\end{equation}
and writing $\ak$ in terms of the field and conjugate momentum operators to write an first order functional differential equation for $\ket \phi(t,\vx)| \alpha \bra$.

\section{Legendre Functions}
We use the following identities found in \cite{DLMF-Legendre}.
\begin{align}
& P^{-\mu}_\lambda(x) \dot{P}^\mu_\lambda(x)-\dot{P}^{-\mu}_\lambda(x)P^\mu_\lambda(x)= \frac{2 \sin(\mu \pi)}{\pi(1-x^2)} \ . \label{eq:P-Wronskian} \\
& P^{i \mu}_\lambda(-x)=\frac{\sin\lp (\lambda+i \mu) \pi\rp}{i \sinh(\pi \mu)} \frac{\Gamma(\lambda+i\mu+1)}{\Gamma(\lambda-i\mu+1)} P^{-i\mu}_\lambda(x) - \frac{\sin(\lambda \pi)}{i \sinh(\pi \mu)} P_\lambda^{i\mu}(x) \ . \label{eq:P-ref} \\
&\lim_{x \ra 1^-} P^\mu_\lambda(x)=\frac{1}{\Gamma(1-\mu)} \lp \frac{2}{1-x} \rp^{\mu/2} \label{eq:1minus-limit}
\end{align}

\bibliographystyle{plain}
\bibliography{bibliography}

\end{document}